\documentstyle[epsfig,color,12pt]{article}

\include{epsf}

\def\cm{{\,\rm cm}}
\def\km{{\,\rm km}}
\def\s{{\,\rm s}}
\def\g{{\,\rm g}}
\def\beq{\begin{equation}}
\def\eeq{\end{equation}}

\begin{document}

 \centerline {\large\bf Regular black hole remnants and graviatoms with de Sitter interior}
\centerline {\large\bf as heavy dark matter candidates probing inhomogeneity of early Universe}

\vskip 0.2in

\centerline{\large\it{Irina~Dymnikova$^{1,2}$} and Maxim
Khlopov$^{3,4,5}$}

\vskip 0.2in

\centerline{\sl $^{1}$ Department of Mathematics and Computer
Science, University of Warmia and Mazury,}

\centerline{\sl S{\l}oneczna 54, 10-710  Olsztyn, Poland; e-mail:
irina@matman.uwm.edu.pl}

\centerline{\sl $^{2}$ A.F. Ioffe Physico-Technical Institute,
Politekhnicheskaja 26, St.Petersburg, 194021 Russia}

\centerline{\sl $^{3}$ APC laboratory 10, rue Alice Domon et L\'eonie Duquet}
\centerline{\sl 75205 Paris Cedex 13, France}

\centerline{\sl $^{4}$ Centre for Cosmoparticle Physics "Cosmion"}
\centerline{\sl $^{5}$ National Research Nuclear University "MEPhI"}
\centerline{\sl (Moscow Engineering physics Institute), 115409 Moscow, Russia}
\begin{abstract}
We address the question of regular primordial black holes with de Sitter interior, their remnants and gravitational
vacuum solitons G-lumps as heavy dark matter candidates providing signatures for inhomogeneity of early Universe,
which is severely constrained by the condition that the contribution of these objects  in the modern density
doesn't exceed the total density of dark matter. Primordial black holes and their remnants seem to be most
elusive among dark matter candidates. However, we reveal a nontrivial property of compact objects with de Sitter
interior to induce proton decay or decay of neutrons in neutron stars. The point is that they can form graviatoms,
binding electrically charged particles. Their observational signatures as dark matter candidates provide also
signatures for inhomogeneity of the early Universe. In graviatoms the cross section of the induced proton decay
is strongly enhanced, what provides the possibility of their experimental searches. We predict proton decay paths
induced by graviatoms in the matter as an observational signature for heavy dark matter searches at the IceCUBE
experiment.

\end{abstract}

{\bf Journal Reference: Int. J. Mod. Phys. D Vol. 24, 1545002
(2015),

special issue "Composite dark matter"}

\section{Introduction}

Cosmological dark matter should consist of stable particles. In
the literature there are considered various alternatives to the
popular candidate of weakly interacting massive point-like
particles \cite{zhitnisky2006,maxim2011}. It includes primordial
black hole which are considered as a reliable source of dark
matter for more than two decades
\cite{mcGibbon1987,copeland1994,carr1994} (for a review
\cite{bergstrom2010,PKrev,KhlopovPBH}).

There are several mechanisms for primordial black hole (PBH)
formation \cite{PKrev,KhlopovPBH}.  PBHs with the mass exceeding
$10^{15}$g should survive to the present time and manifest
themselves as a specific form of dark matter. If the mass of PBHs
is less, than $10^{15}$g, they should evaporate according to the
mechanism of Hawking and their effect in the dark matter is
possible only if there are stable remnants of their evaporation.

Characteristic scales for astrophysical structures originated from
primordial quantum black holes has been considered in
\cite{salvatore2010}. The existence of black hole remnants was
addressed in various aspects since PBHs were predicted in
\cite{N,H}. For example, hybrid inflation can in principle yield
the necessary abundance of primordial black hole remnants for them
to be the primary source of dark matter
\cite{copeland1994,lyth1999}.  Production of primordial black hole
remnants in the early universe can induce a matter-dominated era
before the onset of inflation. Effects of such an epoch on the CMB
power spectrum are discussed in \cite{scardigli2011} where it was
shown that they can be able to explain the quadrupole anomaly of
the CMB power spectrum. Observational scheme for detection of
remnants and their cosmological constraints are discussed in
\cite{nozari2005}.

In the framework of low scale gravity possible signatures of black
hole events in $\it pp$ collisions at the CERN LHC have been
investigated in the hypothesis that black holes do not decay
completely into SM particles \cite{dimopoulos2001} but  leave
behind a (meta-)stable remnant \cite{koch2005}. The production of
black hole remnants has been predicted to occur with the rate of
$10^8$ per year \cite{stoecker2007}. Possible signatures of black
hole events have been considered. The results for the hypothesis
of a stable remnant show  that stable remnant scenarios up to
$M_G\leq ~2~TeV$ can be already probed with the $5~fb^{-1}$ of
integrated luminosity collected by the LHC during the last year of
data taking \cite{bellagamba2012}. Dark matter production at CERN
LHC from black hole remnants  was considered in \cite{nayak2011}.

Let us note that the end-point of black hole evaporation still remains
an open issue \cite{landsberg2006,harris2005,casanova2006}.
In the case of a singular black hole Generalized Uncertainty Principle
requires existence
of a black hole remnant as a Planck size black hole \cite{adler2001,maziashvili2006,banerjee2010}.
 Arguments for remnants based on the nature of the Hawking evaporation and relevant causal domains
 are presented in \cite{ellis2013}. In the Palatini framework a stable remnant with the Planckian
 mass can arise as a geon-like solitonic object supported by the gravitational and electromagnetic
 field \cite{geon}.
On the other hand, no evident symmetry or quantum number was found which would prevent complete
 evaporation \cite{susskind1995}. Analysis of the process of evaporation in the course of formation
  of a black hole shows that the end-point depends essentially on the details of a collapse \cite{depends}.
Character and scale of uncertainty concerning an endpoint of the Hawking evaporation of a singular black hole,
  are clearly evident in the case of a multihorizon spacetime (\cite{me2009} and references therein).
Complete evaporation of a  black hole in de Sitter space creates the problem clearly formulated by Aros \cite{aros}:
In the Schwarzschild-de Sitter space-time the cosmological horizon is not observer-dependent as in the de Sitter space,
but real horizon due to presence of the black hole which breaks the global symmetries involving the radial direction.
A serious doubt concerns a causal structure of space-time: the fate of energy radiated once a black hole disappears
leaving behind the de Sitter space with nothing beyond the cosmological horizon but the de Sitter space itself,
so energy can not be hidden there \cite{aros}. Complete evaporation would create one more serious problem
-- how to evaporate a singularity? \cite{me2009}.

In this paper we consider the regular primordial black hole (RPBH)
remnants and gravitational solitons with de Sitter interior as
heavy dark matter candidates testifying for the inhomogeneity of
early Universe. The pronounced signature for its inhomogeneity is
just  black hole formation. In the case of regular black holes
their evaporation leaves behind stable remnants
\cite{me1996,me2007,me2010}, whose contribution is constrained by
the observed dark matter density.  RPBH and gravitational solitons
with de Sitter interior can arise in a quantum collapse of
primordial fluctuations in the wide range of masses including
those sufficiently small to evaporate to remnants to the end of
inflation \cite{graviatom}. Regular stable remnants with de Sitter
interior should appear in the result of evaporation of a regular
black hole of any origin, provided that the source term satisfies
the condition $T_0^0=T_1^1$ ($p_r=-\rho$)
\cite{me1992,me1996,me2000,me2002}. They represent dark matter
generically related to dark energy through de Sitter interior
where  $p=-\rho$, actually dark ingredients in one drop
\cite{me2007,me2011}. They can form graviatoms with charged
particles whose observational signatures as dark matter candidates
provide also signatures for inhomogeneity of the early Universe.

The paper is organized as follows. In Sect.2 we present compact objects with de Sitter
interior and mechanisms of their production in the early Universe. In Sect.3 we analyze
their observational signatures and in Sect.4 we summarize and discuss the results.

 \section{Regular black holes and solitons with de Sitter interior}

The Einstein equations admit the class of regular solutions asymptotically de Sitter at the center
\cite{me2000,me2002} with a source term specified by $T_0^0=T_1^1$ \cite{me1992}, satisfying the weak energy condition, and identified as
a vacuum dark fluid \cite{me2007}. Solutions from this class, describe a cosmological vacuum dark energy evolving from $T_i^k=\Lambda\delta_i^k$ as $r\rightarrow 0$ to $T_i^k=\lambda\delta_i^k$ as $r\rightarrow\infty$ with $\lambda < \Lambda$ \cite{bdd2003,bd2007,bdg2012},
and compact vacuum objects representing dark matter generically related to dark energy via de Sitter interior \cite{me1992,me2007,me2011}.
In the Universe with non-zero background cosmological constant $\lambda$ (vacuum dark energy) dark matter is represented by regular cosmological
black holes \cite{me1998}, their remnants \cite{me2010}, and  vacuum gravitational solitons G-lumps \cite{me1996,me2002}. These compact objects can originate in the early universe in a way similar to inflation induced (singular) Planck-size black hole
remnants \cite{chen2005} and primordial black holes \cite{nozari2007}.

A regular compact object with de Sitter interior in de Sitter space is described by the line element
\beq
ds^2=g(r)dt^2-\frac{dr^2}{g(r)}-r^2d\Omega^2
   \label{1}
   \eeq
with the metric function \cite{me1998}, asymptotically de Sitter as $r=0$ and asymptotically Schwarzschild-de Sitter as $r\rightarrow\infty$
\beq
g(r)=1-\frac{2G{\cal M}(r)}{r}-\frac{\lambda}{3}r^2;~~~~{\cal M}(r)=4\pi\int^r_0{\rho(x)x^2dx}
   \label{2}
   \eeq
   Spacetime has maximum three horizons \cite{bdd2003}, shown in Fig.1, an internal horizon $r_a$, a black hole horizon $r_b$, and a cosmological horizon $r_c$.
\begin{figure}[htp]
\vspace{8.0mm} \centering \epsfig{file=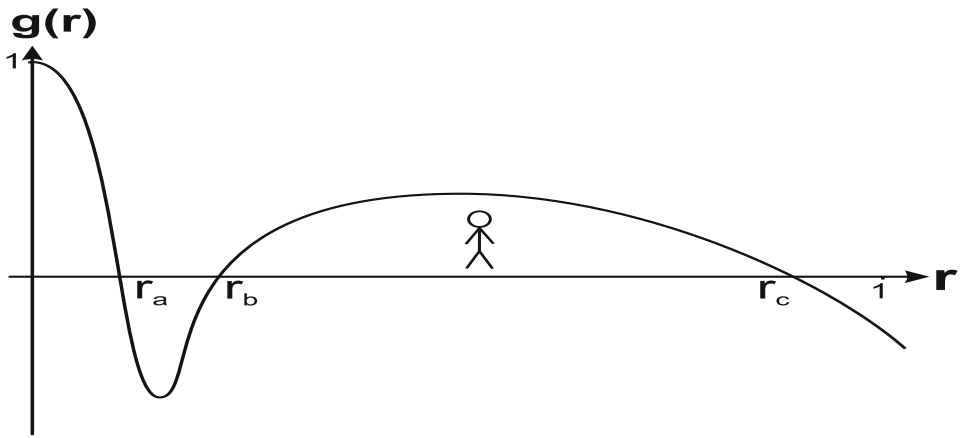,
width=12.7cm,height=5.7cm} \caption{Typical behavior of a metric
function for three-horizons spacetime \cite{bdd2003}.}
\label{Fig.1}
\end{figure}
All three horizons, $r_h$, evaporate with the Gibbons-Hawking
temperature \cite{GH}
   \beq
   k T_h=\frac{\hbar c}{4\pi}|g'(r_h)|
                                                         \label{3}
   \eeq
   Dynamics of horizons evaporation determines
evolution of a black hole \cite{me2010}. In the course of
evaporation a black hole horizon shrinks while horizons $r_a$ and
$r_c$ move outwards, and mass of black hole
$m=4\pi\int_0^{\infty}{\rho(r)r^2dr}$ decreases (for a review
\cite{entropy}). Dependence of the temperature of  a black hole
horizon on the horizon radius, as measured by an observer (shown
in Fig.1) in the R-region between the horizons $r_b$ and $r_c$, is
shown in Fig.2, where we plotted \cite{entropy} $T_b=kT/(\hbar
c/4\pi)$ on the event horizon radius $r=r_b$ normalized to
$r_{\lambda}=\sqrt{3/\lambda}$, for the case
$\sqrt{\Lambda/\lambda}=50$.
\begin{figure}[htp]
\vspace{8.0mm}
\centering \epsfig{file=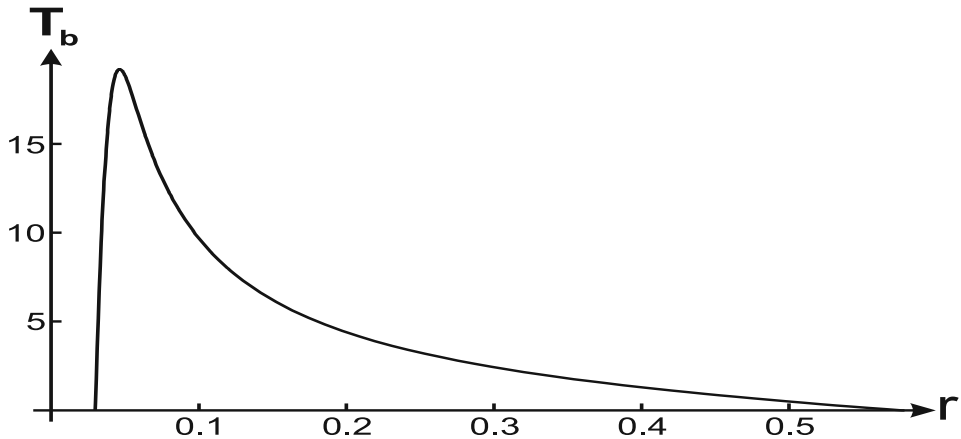, width=12.7cm,height=5.7cm}
\caption{Temperature on a black hole horizon.}
\label{Fig.2}
\end{figure}

 This curve is generic for a regular cosmological  black hole with de Sitter interior and dictated by its two de Sitter asymptotics, in the centre and at infinity \cite{me2010}:
Temperature is zero at the double horizon $r_b=r_c$ (corresponding to the regular modification of the Nariai solution), grows along the branch with the negative specific heat when black hole horizon shrinks, achieves its maximum value $T_{max}$ where specific heat breaks and changes
 sign which testifies for second-order phase transition, after that temperature quickly vanishes at a certain critical value $m_{cr}$ corresponding to the double horizon $r_a=r_b$ with the positive specific heat \cite{me1996,me2010}.

 The fate of a regular black hole
 is thus unambiguous: it leaves behind a thermodynamically stable double-horizon remnant with zero temperature and positive specific heat \cite{me1996,me2010}.

Solutions with $m<m_{cr}$ descrribe gravitational solitons without black hole horizons \cite{me1996} called G-lumps since they hold themselves together by their  own gravity \cite{me2002}.

For the density profile $\rho=\rho_{int}\exp(-r^3/r_0^2r_g)$ \cite{me1992} representing semiclassically vacuum polarization effects \cite{me1996}, critical mass and maximum temperature are given by
\beq
m_{cr}=0.3m_{pl}\sqrt{\rho_{pl}/\rho_{int}}~~~~T_{max}=0.2T_{pl}\sqrt{\rho_{int}/\rho_{pl}}
\label{4}
\eeq
Here $\rho_{int}$ is the density of interior de Sitter vacuum, $r_0^2=3/8\pi G\rho_{int}$, and $r_g=2Gm$.

Most general mechanism of PBH formation in the early Universe involves primordial density inhomogeneities leading to appearance of the overdense regions which can stop expansion and collapse \cite{N,H}. Primordial black holes can form on different stages in the early Universe provided that  mass  contained under its gravitational radius is sufficient for a collapse into a black hole \cite{KhlopovPBH,polnarev2014}.  Regular primordial black holes  with de Sitter interior appear when a quantum collapse of a primordial fluctuation does not lead to formation of a central singularity but stops at achieving, for a certain high density, the state $p=-\rho$  (\cite{graviatom} and references therein).

In the frame of the hypothesis of arising of interior de Sitter vacuum due to symmetry restoration in a collapse at the GUT scale \cite{haifa}, $\rho_{int}=\rho_{GUT}$ and $E_{int}=E_{GUT}\simeq{10^{15}GeV}$, mass of the remnant, its gravitational radius and maximum temperature are  \cite{me1996,me2010}
\beq
 m_{cr}\simeq{0.6\times 10^3~\g},~~ r_g\simeq 10^{-25}\cm,~~
T_{max}\simeq{0.2\times 10^{11}~GeV}
\label{5}
\eeq
In the frame of hypothesis of self-regulation of geometry due to vacuum polarization effects near the Planck scale \cite{W}
or the existence of the limiting curvature of the Planck scale \cite{V}, $E_{int}=E_{Pl}$, mass of the remnant is of order of $m_{Pl}$
and $T_{max}\simeq{0.2 T_{Pl}}$.

Based on the arguments, given above, we consider existence of such stable remnants as an inevitable final stage of evaporation of RPBHs generically related to vacuum dark energy via de Sitter interior.

 Probability of formation of a compact object with mass $m$ and de Sitter interior in a quantum collapse of a quantum fluctuation at a phase transition involving an inflationary vacuum of the scale $E_0$, is estimated as \cite{graviatom}
\beq
D > \exp{\bigg[-4\bigg(\frac{m}{m_{Pl}}\bigg)^{3/4}\bigg(\frac{E_{Pl}}{E_0}\bigg)\bigg]}
\label{6}
\eeq
General constraint on the mass of an object $m$ involves the scale of the interior de Sitter vacuum and reads \cite{graviatom}
\beq
\frac{m}{m_{Pl}} > \bigg(\frac{E_0}{E_{Pl}}\bigg)^4\bigg(\frac{E_{Pl}}{E_{int}}\bigg)^8
\label{7}
 \eeq
 In the case of the GUT scale for both inflationary and interior de Sitter vacuum, $E_{int}=E_{0}=E_{GUT}\simeq{10^{15}GeV}$, it gives the constraint $m > 10^{11}$ g. RPBHs arising in a quantum collapse of primordial quantum fluctuations, have enough time to evaporate producing stable remnants.

 In the case of interior de Sitter vacuum of the Planck scale, $E_{int}=E_{Pl}$, mass range (7) of compact objects emerging from a collapse, admits also G-lumps which have been produced with the bigger probability by virtue of (6).

 One more possibility of production of RPBH and solitons G-lumps from quantum collapse of  quantum fluctuations can be realized at the second inflationary stage predicted by the standard model of particle physics as related to a phase transition at the QCD scale  $E_{0}\simeq{100-200 MeV}$ (\cite{B} and references therein). At this stage quite massive remnants and solitons with $m\leq m_{cr}$ given by (4) could be produced
 although with smaller probability by virtue of (6).

  Another mechanism of production of compact objects with the de Sitter interior can work at the first-order phase transitions. They can form in collisions of true vacuum bubbles arising in a false vacuum background, then a false vacuum of the background scale $E_0$ appears inside an object as $E_{int}$ captured in a collision \cite{KKRS1,KKRS2,DKR}.

  \vskip0.1in

   Regular primordial black holes, their remnants and G-lumps can form graviatoms, i.e., gravitationally bound quantum systems made of such an object as a nucleus and a captured charged particle (\cite{graviatom} and references therein). RPBH and G-lumps arising at the first and second inflationary stages capture available particles. At the end of the first inflationary stage GUT particles can be captured, whose binding energy in graviatoms is comparable with their mass (which makes graviatoms stable), and survive to the present epokh as constituents of graviatoms   \cite{graviatom}. Electromagnetic radiation of the graviatom can bear information about a fundamental symmetry scale of its interior de Sitter vacuum  and serve as its observational signature. Characteristic frequency of the oscillatory type radiation of graviatom is $\hbar\omega=0.678(\hbar c/r_{int})$ where $r_{int}$ is the de Sitter radius $r_{int}^2=3c^2/8\pi G\rho_{int}$. It gives $\hbar\omega=0.678\times 10^{11}$GeV$(E_{int}/E_{GUT})^2$ \cite{graviatom}. Current experiments allow detection of photons up to $10^{11.5}$ GeV \cite{troitsky} so that radiation falls within the range of observational possibilities for graviatoms with the GUT scale de Sitter interior.

\section{Observational signatures}

The concentration of  stable remnants of  RPBHs is restricted by the condition that their modern contribution into the total density doesn't exceed the total density of dark matter.

Consider, for simplicity, the case, when RPBH of mass $m\ge m_{cr}$ is formed in the early Universe on the stage of relativistic expansion with the equation of state $p=\epsilon/3$ at the time $t_i(m)=m/m_{pl}^2$ and evaporates on radiation dominance (RD) stage at $t_e(m)=m^3/m_{pl}^4$.
If the fraction of the total cosmological density in the period of PBH formation was $\beta(m)$, this contribution grew at the RD stage inversely proportional to the scale factor $\propto (t/t_i)^{1/2}$ and reached $\alpha(m)=(t_e(m)/t_i(m))^{1/2} \beta(m)=(m/m_{pl})\beta(m)$ in the period of evaporation.

Since the evaporation of PBHs with mass $m\ge m_{cr}$ ends by formation of remnant with mass $m_{cr}$, these remnants contribute into the total density after evaporation as $$\beta(m,m_{cr})=\frac{m_{cr}}{m_{pl}}\beta(m).$$
For PBHs, evaporating before the end RD stage at $t_{rd}$, the contribution of remnants into the total density grows $\propto (t/t_e)^{1/2}$ and becomes equal to $$\alpha_{rd}=\left(\frac{t_{rd}}{t_e(m)}\right)^{1/2}\frac{m_{cr}}{m_{pl}}\beta(m)=\left(\frac{t_{rd}}{t_{pl}}\right)^{1/2}\left(\frac{m_{pl}}{m}\right)^{1/2}\frac{m_{cr}}{m}\beta(m)$$ at $t=t_{rd}$.

The condition $\alpha_{rd} \le 1$ that this contribution doesn't exceed the total dark matter density in the period of matter dominance after the end of RD stage at $t=t_{rd}$ results in the following constraint on the fraction $\beta(m)$ of the total density, corresponding to PBHs with mass $m\ge m_{cr}$ in the period of their formation
\beq
\beta(m)\le \left(\frac{t_{pl}}{t_{rd}}\right)^{1/2}\left(\frac{m}{m_{pl}}\right)^{1/2}\frac{m}{m_{cr}}.
\label{8}
\eeq

It provides a strong nontrivial constraint on $\beta(m)$ for PBHs with mass $m < 10^9 g$, which evaporate before the era of Big Bang Nucleosynthesis and for which the model independent constraint from the observed entropy of the Universe is rather weak.

The constraint Eq.(\ref{8}) is the strongest, when the mass of evaporating PBHs is close to its minimal possible value, corresponding to the mass of the stable remnant. In this case the restriction Eq.(\ref{8}) reads as
\beq
\beta(m_{cr})\le \left(\frac{t_{pl}}{t_{rd}}\right)^{1/2}\left(\frac{m_{cr}}{m_{pl}}\right)^{1/2} \approx 10^{-24},
\label{9}
\eeq
where the numerical estimation is given for $m_{cr}\simeq{0.6\times 10^3~\g}$, $t_{pl}=5 \cdot 10^{-44}\s$, $t_{rd}=5 \cdot 10^{12}\s$ and $m_{pl}\simeq{2\times 10^{-5}~\g}$.

It should be noted that the constraint Eq.(\ref{8}) can become weaker with the account for possible existence of early dust-like stages, at which the contribution of PBHs into the total density doesn't grow. However, on the other hand, the probability of PBH formation at such stages can grow and in any case, even weakened, the constraint Eq.(\ref{8}) still remains much stronger than the one from the entropy of the Universe.
One should also take into account that stronger constraints on miniPBHs existing in the literature assume the existence of stable hypothetical particles, like moduli \cite{lemoine} or gravitino \cite{BGK}, while
the constraint obtained in the present work is based on the existence of stable remnants only and is thus independent of particle physics models.

The case of strict equality in the Eq.(\ref{8}) corresponds to the dominance of remnants in the dark matter density.
This form of dark matter may be the most elusive in the list of dark matter candidates, owing to a very small cross section of their interaction with matter and between themselves.

Indeed, geometrical cross section of these objects is of the order of $\sigma_g\simeq \pi r_g^2 \simeq 3 \cdot 10^{-50}\cm^2$. This value can be compared with the sensitivity of direct WIMP searches in the underground detectors, which is expected in the planned most sensitive XENON1T detector to be $\sigma_s\simeq 3 \cdot 10^{-47}\cm^2$ for spin independent interaction of WIMPs with the mass about 50GeV. However, the number density of remnants in vicinity of Solar system should be by 29 orders of magnitude less, than the number density of these WIMPs, what makes impossible their direct experimental searches.

Taking formally accretion radius for the considered remnants as $r_a=2Gm_{cr}/v^2$, one obtains $r_a \simeq 10^{-19}\cm$ for velocity in halo $v\simeq 300\km/\s$, which is hundreds thousand times smaller, than the size of nuclei. Therefore, penetration of remnants in Solar matter with the number density $n \sim 10^{24} \cm^{-3}$ cannot lead to a significant effect of accretion, since $n \sigma_a R_\bigodot \approx 2 \cdot 10^{-3}$, where $\sigma_a= \pi r_a^2$ is the accretion cross section and $R_\bigodot \approx 7 \cdot 10^{10} \cm$ is the Solar radius. Even in neutron stars, with particle number density $n_{ns} \sim 10^{38} cm^{-3}$ and radius $R_{ns} \sim 10^6 \cm$, in which $ n_{ns} \sigma_a R_{ns} \approx 3 \cdot 10^6$, the increase of remnant's mass due to accretion $\Delta m = \rho_{ns} \sigma_a R_{ns} \approx 3 \cdot 10^{-18} \g$ is much less than the mass of the remnant, so that the corresponding decrease of velocity is negligible, what prevents their capture.

The situation can change drastically, if we take into account that the regular remnant
is a compact object with the GUT false vacuum interior, in which baryon and lepton numbers
are not conserved. It may lead to gravitational capture of proton that induces proton decay,
similar to Callan-Rubakov effect in the case of GUT magnetic monopoles \cite{callan,rubakov}.
However, on the contrary to the case of magnetic monopoles due to much larger mass of remnants,
their number density in the vicinity of the Solar system is too small to have a hope to observe this effect.
Indeed, if we take the
 the cross section of induced proton decay due to its gravitational capture by remnants
 as $ \sigma_a= \pi r_a^2 \approx 3\cdot 10^{-38} \cm^2$ and assume that remnants saturate
 the dark matter density with local number density $n_r \approx 10^{-26} \cm^{-3}$ in the ton
 of matter of an underground detector one should expect one event per $10^{26} \s \approx 3\cdot 10^{18} years$.

 As the extreme case, which needs special study in the framework of e.g. bag models, the very penetration of
 remnant inside the nucleon may induce its decay. But even if the corresponding cross section is determined
 by the geometrical size of nucleon ($\sigma_i \sim 10^{-26} \cm^2$) one can expect no more than one event
 per $10^7$ years in one ton of an underground detector. In the matter of a $1 \km^3$ detector, like IceCUBE,
 there should be up to 300 events per year. However, the problem of distinguishing such events,
 in which only 1 GeV energy is released, is crucial in this case.

 If induced nucleon decay can take place during penetration of remnant through the neutron star, the remnant is not captured but crosses the star leaving a pin-hole of decayed neutrons in it. This picture is essentially different from the case of much more massive PBHs, whose capture by neutron stars was constrained by \cite{capela}. For the induced decay $n \rightarrow \pi^0 + \nu$ about half of the neutron rest energy is released in a nuclear size area owing to $\pi^0$ interaction with the neutron star matter, since the timescale of this interaction is much less, than the lifetime of $\pi^0$, $\tau_{0}$,  $t_i \sim (n_{ns}\sigma_{\pi n} v)^{-1} \sim 10^{-22}\s \ll \tau_{0}$. The corresponding pressure can exceed the gravitational pressure and the matter of the neutron star can be ejected in the form of a very narrow collimated jet. Such jets, corresponding to the crossing of remnants through a neutron star can appear at the rate $n_r \pi R_{ns} v \sim 10^{-6} \s^{-1}$, i.e. about 30 per year per neutron star.

 Graviatoms with de Sitter interior can contribute to composite dark matter whose
 observational signatures are related  to electromagnetic radiation which bears information
about their interior structure, in particular about the
fundamental symmetry scale responsible for de Sitter vacuum
interior \cite{graviatom}. Current experiments allow detection of
photons up to $10^{11.5}$~ GeV (\cite{kalashev} and references
therein). Present observational possibilities prefer thus
graviatoms with the GUT scale interior although those with the
Planck scale interior can exist in principle, and probabilities of
their production in a collapse are bigger, but their typical
frequency, $\hbar\omega\simeq{0.7\times 10^{19}}$ GeV is far from
the today observational range.

For the charged remnants one can expect also much stronger effect
of falling down its center and correspondingly much stronger
effect of induced proton decay. If the corresponding cross section
is of the same order, as in the case of magnetic monopole
$\sigma_i \sim 10^{-28} \cm^2$, one should expect about 3 events
per year of 1 GeV energy release due to slow penetrating massive
particles in IceCUBE. Search for this effect needs special study
in the analysis of the IceCUBE data.

Let us emphasize that graviatoms are bound states of a remnant
with false vacuum interior and an electrically charged particle,
captured by it. Therefore one may expect that the remnant
component of graviatom can induce nucleon decay, while the charged
component provides the enhancement of the corresponding cross
section. It makes the searchs for graviatoms challenging for
IceCUBE experiment.

\section{Conclusions}

Regular primordial black holes and gravitational solitons with de Sitter vacuum interior can arise in the early Universe from primordial inhomogeneities in assumption that quantum collapse of primordial overdense regions stops at a certain stage when the pressure
becomes $p=-\rho$ as a result of symmetry restoration or existence of the limiting density/curvature scale. Evaporation of RPBH leaves behind stable remnants with zero temperature and positive specific heat. They can be considered as heavy dark matter candidates whose observational signatures can serve as signatures for inhomogeneity of the early Universe. RPBH, their remnants and solitons can form graviatoms by capturing available charged particles. Radiation of graviatoms can bear information on the scale of the interior de Sitter vacuum. In the case of the GUT scale interior, radiation of the oscillatory type falls within the ultra-high energy range available in principle for today observational possibilities.

The mass of RPBH remnants is about $10^2-10^3$ kg, what makes strongly suppressed their gravitational cross section, so they are elusive dark matter candidates, although not as elusive as Planck mass remnants.  We have shown that for the remnants with GUT vacuum interior induced proton or neutron decay is possible. The corresponding cross section is strongly enhanced for electrically charged graviatoms, making their search as heavy dark matter candidates challenging in the IceCUBE experiment.

\section*{Acknowledgment}

The work by I.D. was partly supported by the National Science Center of Poland through the
grant 5828/B/H03/2011/40. The work by M.Kh.
on initial cosmological conditions was supported by the Ministry of Education and Science of Russian Federation,
project 3.472.2014/K  and his work on the forms of dark matter was supported by grant RFBR 14-22-03048.


\begin{thebibliography}{99}

\bibitem{zhitnisky2006}
A. Zhitnisky, Phys. Rev. {\bf D 74} (2006) 043515.


\bibitem{maxim2011}
M.Yu. Khlopov, Mod. Phys. Lett. {\bf A 26} (2011) 2823.


\bibitem{mcGibbon1987}
J.H. MacGibbon, Nature {\bf 329} (1987) 308.

\bibitem{copeland1994}
E. J. Copeland et al, Phys. Rev. {\bf D 49} (1994) 6410.

\bibitem{carr1994}
B. J. Carr, Proc. of 22nd Texas Symp., ECONFCO {\bf 41213}
(2004) 0204; astro-ph/0504034.

\bibitem{bergstrom2010}
L. Bergstrom, the plenary talk at the "Invisible Universe
International Conference", Paris, France, June-July 2009,
 in \emph{AIP Proceedings Series} (2010).

\bibitem{PKrev}
  A.G. Polnarev, M.~Y.~Khlopov, Sov.Phys.Usp. {\bf 28} (1985) 213.

\bibitem{KhlopovPBH}
  M.~Y.~Khlopov,
  Res.\ Astron.\ Astrophys.\  {\bf 10} (2010) 495
  [arXiv:0801.0116 [astro-ph]].


\bibitem{salvatore2010}
S. Capozziello, G. Cristofano, M. De Laurentis, Eur. Phys.J. {\bf
C 69} (2010) 293.

\bibitem{N}
Ya.B. Zeldovich and I.D. Novikov, Sov. Astron. {\bf 10} (1967) 602.

\bibitem{H}
S.W. Hawking, Mon. Not. R. Astron. Soc. {\bf 152} (1971) 75.

\bibitem{lyth1999}
D. H. Lyth and A. Riotto, Phys. Rep. {\bf 314} (1999) 1.

\bibitem{scardigli2011}
F. Scardigli, C. Gruber and P. Chen, Phys. Rev. {\bf D 83} (2011) 063507.

\bibitem{nozari2005}
K. Nozari and S.H. Mehdipour, Mod. Phys. Lett. {\bf A 20} (2005) 2937.

\bibitem{dimopoulos2001}
S. Dimopoulos and G. Landsberg, Phys. Rev. Lett. {\bf 87} (2001) 161602.

\bibitem{koch2005}
B. Koch, M. Bleicher and S. Hossenfelder, JHEP {\bf 0510} (2005) 053.

\bibitem{stoecker2007}
H. Stoecker, Int. J. Mod. Phys. {\bf D 16} (2007) 185.

\bibitem{bellagamba2012}
L. Bellagamba, R. Casadio, R. Di Sipio, V. Viventi, arXiv:1201.3208 [hep-ph] 26 Mar 2012.

\bibitem{nayak2011}
G.C. Nayak, Phys. Part. Nucl. Lett. {\bf 8} (2011) 337.


\bibitem{landsberg2006}
G. Landsberg, J. Phys. {\bf G 32} (2006) R337.

\bibitem{harris2005}
C.M. Harris, M.J. Palmer, M.A. Parker, P. Richardson, A. Sabetfakhri, B.R. Webber, JHEP {\bf 0505} (2005) 053.

\bibitem{casanova2006}
A. Casanova and E. Spallucci, Class. Quant. Grav. {\bf 23} (2006) R45.

\bibitem{adler2001}
R.J. Adler, P. Chen and D.I. Santiago, Gen. Rel. Grav. {\bf 33} (2001) 2101.

\bibitem{maziashvili2006}
M. Maziashvili, Phys. Lett. {\bf B 635} (2006) 232.

\bibitem{banerjee2010}
K. Banerjee and S. Ghosh, Phys. Lett. {\bf B 688} (2010) 224.

\bibitem{ellis2013}
George F.R. Ellis, http://arxiv.org/abs/1310.4771 (2013).

\bibitem{geon}
F.S.N. Lobo, G.J. Olmo and D. Rubiera-Garcia, JCAP {\bf 07} (2013) 011.

\bibitem{susskind1995}
L. Susskind, J. Math. Phys. {\bf 36}, 6377 (1995).

\bibitem{depends}
H. Kawai, Y. Matsuo, Y. Yokokura, http://arxiv.org/abs/1302.4733 (2013).


\bibitem{me2009}
I. Dymnikova, the invited talk at the "Invisible Universe
International Conference", Paris, France, June-July 2009,
in \emph{AIP Proceedings Series} (2010).

\bibitem{aros}
R. Aros, Phys. Rev. {\bf D 77} (2008) 104013.

\bibitem{me1996}
I. Dymnikova, Int. J. Mod. Phys. {\bf  D 5}, 529 (1996).

\bibitem{me2007}
I. Dymnikova and E. Galaktionov, Phys. Lett. {\bf B 645} (2007) 358.

\bibitem{me2010}
I. Dymnikova and M. Korpusik, Phys. Lett. {\bf B 685}, 12 (2010).

\bibitem{graviatom}
I. Dymnikova and M. Fil'chenkov, AHEP {\bf 13} (2013) Article ID 746894.

\bibitem{me1992}
I.G. Dymnikova, Gen. Rel. Grav. {\bf 24} (1992) 235.


\bibitem{me2000}
I.G. Dymnikova, Phys. Lett. {\bf B 472} (2000) 33.

\bibitem{me2002}
I. Dymnikova,  Class.Quant.Grav. {\bf 19}, 725 (2002).

\bibitem{me2011}
I. Dymnikova and E. Galaktionov, Central European Journal of Physics {\bf 9} (2011)
644.

\bibitem{bdd2003}
K.A. Bronnikov, A. Dobosz and I. Dymnikova, Class. Quant. Grav. {\bf 20} (2003) 3797.

\bibitem{bd2007}
K.A. Bronnikov and I. Dymnikova, Class. Quant. Grav. {\bf 24} (2007) 5803.

\bibitem{bdg2012}
K.A. Bronnikov, I. Dymnikova and E.V. Galaktionov, Class. Quant. Grav. {\bf 29} (2012) 095025.





\bibitem{me1998}
I.Dymnikova and B. Soltysek, Gen. Rel. Grav. {\bf 30} (1998) 1775.

\bibitem{chen2005}
P. Chen, New Astron. Rev. {\bf 49} (2005) 233.

\bibitem{nozari2007}
K. Nozari, Astropart. Phys. {\bf 27} (2007) 169.



\bibitem{GH}
G.W. Gibbons and S.W. Hawking, Phys. Rev. {\bf D 15} (1977) 2738.

\bibitem{entropy}
I. Dymnikova and M. Korpusik, Entropy {\bf 13} (2011) 1967.

\bibitem{polnarev2014}
T. Nakama, T. Harada, A. Polnarev and J. Yokoyama, JCAP {\bf 01} (2014) 037.

\bibitem{haifa}
I. Dymnikova, in: Internal structure of black holes and spacetime singularities, Eds. M. Burko and A. Ori, Bristol <in-t of Physics qnd the Israel Physical Society (1997) 422.

\bibitem{W}
E. Poisson and W. Israel, Class. Quant. Grav. {\bf 5} (1988) L201.


 \bibitem{V}
 V.P. Frolov, M.A. Markov and V.F. Mukhanov, Phys. Rev. {\bf 41} (1990) 383.






\bibitem{B}
D. Boyanovski, H.J. de Vega and D.J. Schwarz, Ann. Rev. Nucl. Part. Sci. {\bf 56} (2006) 441.

\bibitem{KKRS1}R.V. Konoplich, S.G. Rubin, A.S. Sakharov, M.Yu. Khlopov,
Phys.Atom.Nucl. {\bf 62} (1999) 1593.

\bibitem{KKRS2}R.V. Konoplich, S.G. Rubin, A.S. Sakharov, M.Yu. Khlopov,
Gravitation \& Cosmology {\bf 6} (2000) 153.

\bibitem{DKR}I.Dymnikova, M.Yu. Khlopov, L. Koziel and S. G. Rubin.
Gravitation \& Cosmology {\bf 6} (2000) 311; e-Print ArXive: hep-th/0010120.

\bibitem{troitsky}
O.E. Kalashev, G.I. Rubtsov, and S.V. Troitsky, Phys. Rev. {\bf D 80} (2009) Article ID 103006.

\bibitem{lemoine}
  M. Lemoine, Phys. Lett. B {\bf 481} (2000) 333, arXiv:hep-ph/0001238.

\bibitem{BGK} M.Yu.Khlopov, A.Barrau, and J.Grain,
Class. Quantum  Grav. {\bf 23} (2006) 1875, arXiv:astro-ph/0406621.





\bibitem{callan}
C.G. Callan, Nucl. Phys. {\bf B212} (1983) 391.
\bibitem{rubakov}
V.A. Rubakov, Nucl. Phys. {\bf B203} (1982) 311.
\bibitem{capela}
F. Capela, M. Pshirkov, P. Tinyakov,
Phys. Rev. {\bf D 87} (2013) 123524.
\bibitem{kalashev}
O.E. Kalashev, G.I. Rubtsov, and S.V. Troitsky, Phys. Rev. {\bf D
80} (2009) 103006.

\end{thebibliography}
\end{document}